\documentclass[aps,prd,twocolumn,superscriptaddress,nofootinbib]{revtex4-2}

\usepackage[colorlinks=true,citecolor=red,urlcolor=blue]{hyperref}
\usepackage{amssymb}
\usepackage{amsmath,color}
\usepackage{graphicx}
\usepackage{bbm}
\usepackage{verbatim}
\usepackage[T1]{fontenc}
\usepackage[utf8]{inputenc}
\usepackage[normalem]{ulem}

\usepackage{url}
\usepackage{xcolor}
\usepackage{caption}
\usepackage{subcaption}
\usepackage{orcidlink}

\usepackage{diagbox}
\usepackage{color}
\usepackage{multirow}
\usepackage{hhline}
\usepackage{tabularx}

\begin{document}

\title{Enhancing Early Detection and Localization of Gravitational Waves via Eccentricity-Induced Higher Harmonic Modes with 2G Detector Networks}

\author{Tao Yang\orcidlink{0000-0002-2161-0495}}
\email[]{yangtao@whu.edu.cn}
\affiliation{School of Physics and Technology, Wuhan University, Wuhan 430072, China}

\author{Rong-Gen Cai\orcidlink{0000-0002-3539-7103}} 
\email[]{cairg@itp.ac.cn}
\affiliation{CAS Key Laboratory of Theoretical Physics, Institute of Theoretical Physics, Chinese Academy of Sciences, Beijing 100190, China}
\affiliation{School of Physical Science and Technology, Ningbo University, Ningbo, 315211, China}
\affiliation{School of Fundamental Physics and Mathematical Sciences, Hangzhou Institute for Advanced Study (HIAS), University of Chinese Academy of Sciences, Hangzhou 310024, China}

\author{Zhoujian Cao\orcidlink{0000-0002-1932-7295}}
\email[]{zjcao@bnu.edu.cn}
\affiliation{Department of Astronomy, Beijing Normal University, Beijing 100875, China}
\affiliation{School of Fundamental Physics and Mathematical Sciences, Hangzhou Institute for Advanced Study (HIAS), University of Chinese Academy of Sciences, Hangzhou 310024, China}

\author{Hyung Mok Lee\orcidlink{0000-0003-4412-7161}}
\email[]{hmlee@snu.ac.kr}
\affiliation{Center for the Gravitational-Wave Universe, Astronomy Research Center, Seoul National University, 1 Gwanak-ro, Gwanak-gu, Seoul 08826, Korea}

\date{\today}

\begin{abstract}

Early detection and localization of gravitational waves (GWs) are essential for identifying electromagnetic (EM) counterparts, playing a key role in multi-messenger astronomy. 
However, second-generation (2G) ground-based detectors are most sensitive to frequencies of tens to hundreds of hertz, limiting the in-band duration of GW signals to $\mathcal{O}(0.1)$ to several tens of seconds. 
This constraint hinders early-warning capabilities and early localization.
We present the first theoretical study on how eccentricity-induced higher harmonic modes, which enters the detector band significantly earlier than the dominant mode, enhance early detection and localization in a 2G detector network. 
By decomposing each harmonic mode in the frequency domain and tracking their sequential entry into the detector band, we analyze the evolution of the average signal-to-noise ratios (SNRs) and localization accuracy as functions of time-to-merger.
For a GW170817-like BNS, an eccentricity of $e_0=0.4$ at 10 Hz allows the signal to reach SNR 4 and the detection threshold of SNR 8 approximately 12 and 5 minutes before merger, respectively—gains of 4.5 and 1.5 minutes over the circular case. 
Localization within $1000 \, (100)\,\rm deg^2$ is achievable 5 (1) minutes before merger, improving by 2 minutes (15 seconds).
Our results highlight the potential of eccentricity-induced higher harmonics in improving early warnings and localization, particularly for BNS mergers, enhancing the prospects for multi-messenger astronomy.

\end{abstract}

\maketitle

\section{Introduction}
Early detection and precise localization of gravitational waves (GWs) are pivotal in advancing multi-messenger GW astronomy and enhancing their applications in cosmology, astrophysics, and fundamental physics~\cite{LIGOScientific:2017ync,LIGOScientific:2017zic,LIGOScientific:2017adf,Creminelli:2017sry,Ezquiaga:2017ekz,Baker:2017hug,Mooley:2018qfh,Hotokezaka:2018dfi,Dietrich:2020efo,LIGOScientific:2019gag,Bian:2025ifp}.
In particular, low-latency or even negative-latency GW alerts, along with preparatory and follow-up observations of electromagnetic (EM) counterparts, would greatly benefit from earlier pre-merger detection and localization.
The feasibility of pre-merger detection and localization depends primarily on the in-band signal duration and the detector network configuration. 
For the second-generation (2G) detectors LIGO-Virgo-KAGRA (LVK) at current sensitivity levels, the in-band time is typically limited to $\mathcal{O}(0.1)$ to several tens of seconds, assuming an initial frequency of 30 Hz. 
This duration is often insufficient to accommodate the slew times of many EM telescopes, which range from 30 to 60 seconds~\cite{Kapadia:2020kss}. 
Enhancing the prospects of detecting EM counterparts and advancing GW multi-messenger astronomy with 2G detectors thus requires new methodologies and technological innovations.

Orbital eccentricity is a crucial feature for distinguishing between different binary black hole (BBH) formation scenarios~\cite{Nishizawa:2016jji,Nishizawa:2016eza,Breivik:2016ddj,Zevin:2021rtf}. 
Over the last decade, various studies have explored eccentric gravitational waves (GWs) from multiple perspectives, 
including the template~\cite{Yunes:2009yz,Huerta:2014eca,Tanay:2016zog,Huerta:2016rwp,Cao:2017ndf,Huerta:2017kez,Moore:2019xkm,Ramos-Buades:2021adz,Liu:2023ldr,Arredondo:2024nsl,Gamboa:2024imd}, 
detection~\cite{Lower:2018seu,LIGOScientific:2019dag,Romero-Shaw:2019itr,Nitz:2019spj,Wu:2020zwr,Romero-Shaw:2020thy,Gayathri:2020coq,Iglesias:2022xfc,LIGOScientific:2023lpe,Gupte:2024jfe},
and parameter estimation~\cite{Favata:2013rwa,Sun:2015bva,Gondan:2017hbp,Saini:2022igm,Narayan:2023vhm,Yang:2022fgp,GilChoi:2022waq}.
The improvement of localization of GWs from eccentricity has been investigated by~\cite{Mikoczi:2012qy,Kyutoku:2013mwa,Ma:2017bux,Pan:2019anf}. 
Our previous studies~\cite{Yang:2022tig,Yang:2022fgp,Yang:2022iwn} shows that the eccentricity of long-inspiraling compact binaries can enhance distance estimation and source localization accuracy by several orders of magnitude when observed with space-based decihertz observatories.

Various dynamical formation mechanisms for compact binaries involving black holes and neutron stars have been proposed to explain their potential eccentricities~\cite{Rodriguez:2017pec,Samsing:2017xmd,Samsing:2017oij,Samsing:2018ykz,Wen:2002km,Pratten:2020fqn,OLeary:2008myb,Lee:2009ca}. However, the rate of detecting eccentric BBH and BNS mergers in the LVK band remains highly uncertain with current knowledge.
Several studies~\cite{Samsing:2017xmd,Samsing:2017rat} suggest that approximately $\mathcal{O}(5\%)$ of dynamical mergers occurring in globular clusters may exhibit eccentricities greater than 0.1 at a gravitational wave frequency of 10 Hz. 
Moreover, the recent report of a high eccentricity ($e_0 \sim 0.6$) for GW190521~\cite{Gayathri:2020coq} indicates the potential for observing more highly eccentric GW events in the LVK band in the future.

The enhancement of early warning and parameter estimation of GWs by incorporating higher multipole modes from asymmetric mass ratios has been explored in~\cite{Kapadia:2020kss,Singh:2020lwx,Islam:2021zee}.
Intriguingly, our recent study~\cite{Yang:2023zxk} demonstrates that eccentricity-induced higher harmonic modes allow the third-generation (3G) detector network ET+2CE to detect and localize typical sources even before the dominant mode enters the band, enabling the earliest possible warning and maximizing preparation time for observing EM counterparts.
However, given the limited in-band duration of GWs in 2G detectors, improving early warning and localization in 2G networks is both more pressing and challenging.

In this paper, through the decomposition and extraction of each harmonic mode in the frequency domain, we present the first theoretical and comprehensive study on the enhancement of early detection and localization from eccentricity-induced higher harmonic modes in a 2G detector network—Advanced LIGO, Advanced Virgo, KAGRA, and LIGO-India (HLVKI).
The higher harmonic modes enter the detector band much earlier than the dominant mode, allowing for early SNR accumulation and localization estimation.
Our focus lies on the evolution of SNR and localization accuracy as functions of time-to-merger, as well as the time gained for pre-merger detection and localization enabled by eccentricity-induced higher harmonic modes.
By simulating typical compact binaries and tracking the sequential entry of each harmonic mode into the detector band, we quantify the extent to which eccentric cases can achieve earlier accumulation of a given SNR and localization accuracy compared to non-eccentric cases.

\section{Methodology}
In this paper, as we need the frequency decomposition of higher harmonic modes and focus on the pre-merger SNR and localization, 
we adopt the non-spinning, inspiral-only, frequency-domain EccentricFD waveform approximant provided by the {\sc LALSuite} software package~\cite{lalsuite}.
The waveform can be expressed as the sum of  harmonics~\cite{Huerta:2014eca}: 
\begin{equation}
\tilde{h}(f) = \sum_{\ell=1}^{10} \tilde{h}_\ell(f) \,,
\label{eq:hf}
\end{equation}
where $\tilde{h}_\ell(f)$ represents the $\ell$-th harmonic, which has a frequency $\ell$ times the orbital frequency $F$.
The detailed definitions and expressions in $\tilde{h}_{\ell}(f)$ can be found in~\cite{Yunes:2009yz, Huerta:2014eca}.
The eccentric waveforms are then generated using {\sc PyCBC}~\cite{alex_nitz_2022_6912865}.  
When $e_0 = 0$, only the $\ell=2$ mode is present in the waveform, reducing it to the TaylorF2 waveform model. 
In contrast, for eccentric cases, additional subdominant higher modes also contribute to the waveform.

Since the higher modes (e.g., $\ell=10$) can remain in-band (assuming an initial frequency of 10 Hz) for approximately 20 hours before merger, we account for the Earth’s rotational effect throughout this paper. 
Consequently, the time- (and hence frequency-) dependent beam pattern functions $F_{+,\times}$ of the detector for different higher modes must be calculated separately. We modify {\sc LALSuite} to extract each harmonic's polarization $\tilde{h}_{\ell +,\times}(f)$ individually.
The positions and orientations of the arms for the 2G detector network HLVKI are detailed in Table I of~\cite{Pan:2019anf}. 
The time-dependent $F_{+,\times}(t)$ are constructed using a geocentric coordinate system.
For the eccentric cases, we follow~\cite{Yunes:2009yz} (see equation (4.24)) to numerically solve the phase evolution of the eccentric orbits and obtain $t(f_{\rm ref})$, with the reference frequency corresponding to that of the dominant $\ell=2$ mode. 
We then derive the frequency-dependent $F_{+,\times}(f_{\rm ref})$ based on the reference frequency and for each $\ell$-th mode, the corresponding $F_{+,\times}(f)$ is given by $F_{+,\times}(2f/\ell)$. 
After applying the corresponding antenna pattern function to each mode, the final waveform is obtained by summing all the modes.


We select a GW170817-like binary neutron star (BNS) and a GW150914-like BBH for simulating typical compact binaries
\footnote{In this paper, we do not present results for a typical neutron star–black hole (NSBH) binary, as it does not provide additional conclusions. 
For such unequal-mass systems, the multipolar higher modes become more prominent, and we leave the exploration of these systems for future research.}. 
The component masses ($m_1,m_2$), redshift ($z$), and luminosity distance ($d_L$) are set to match the median values of the actual events\footnote{\url{https://gwosc.org/eventapi/html/GWTC/}}.
To account for the varying sky positions and orientations of the systems, we generate 1000 random sets of angular parameters ($P_{\rm ang}$), including the inclination angle ($\iota$), sky location ($\theta, \phi$), polarization angle ($\psi$), and longitude of the ascending nodes axis ($\beta$), sampled from a uniform and isotropic distribution.
Given the validated use of this eccentric waveform for initial eccentricities up to 0.4~\cite{Huerta:2014eca}, 
we consider five discrete initial eccentricities: $e_0=0,~0.05,~0.1,~0.2, \text{ and } 0.4$ at the initial frequency of the 2G detectors, $f_0=10$ Hz.
In total, we simulate 10,000 events.
We present the averaged results across the 1000 samples of $P_{\rm ang}$ for each typical binary at a given eccentricity.

Since we need to analyze 10,000 events at various specific times to calculate the corresponding SNR and perform parameter estimation for localization, it is impractical to use stochastic sampling algorithms such as the Markov Chain Monte Carlo (MCMC) method.
We adopt the approach of~\cite{Yang:2022tig} and utilize the Fisher matrix technique for gravitational waves~\cite{Cutler:1994ys} to estimate the uncertainties and covariances of the waveform parameters.
The Fisher matrix is defined as 
\begin{equation}
\Gamma_{ij}=(\partial_i h,\partial_j h)\,,
\label{eq: FM}
\end{equation}
where $\partial_i h=\partial h/\partial P_i$ and $P_i$ represents a parameter in the waveform.
The network Fisher matrix is computed as the sum of $\Gamma_{ij}$ from each detector.
The inner product is defined as 
\begin{equation}
(a,b)=4\int_{f_{\rm min}}^{f_{\rm max}}\frac{\tilde{a}^*(f)\tilde{b}(f)+\tilde{b}^*(f)\tilde{a}(f)}{2 S_n(f)}df\,,
\label{eq:inner}
\end{equation}
where $f_{\rm min}$ and $f_{\rm max}$ denote the low- and max-frequency cutoffs.
We set $f_{\rm min} = 10$ Hz as the initial frequency of the 2G detectors, while $f_{\rm max}$ is determined based on the time to merger at which the calculation is performed. 
Note that each mode in Eq.~(\ref{eq:hf}) must be truncated at its respective $f_{\rm max}$.
The noise power spectral densities $S_n(f)$ are the designed sensitivities for advanced LIGO, Virgo, and KAGRA.
Then the SNR is 
\begin{equation}
\rho=(h,h) \,.
\label{eq:SNR}
\end{equation}
The network SNR is calculated as the quadrature sum of the SNRs from each detector.
The sky localization error is given by~\cite{Cutler:1997ta}
\begin{equation}
\Delta \Omega=2\pi |\sin(\theta)|\sqrt{C_{\theta\theta}C_{\phi\phi}-C_{\theta\phi}^2} \,,
\label{eq:loc}
\end{equation}
with the covariance matrix of the parameters as $C_{ij}=(\Gamma^{-1})_{ij}$. 
We need to note the limitations of the Fisher matrix in the parameter estimation of GWs, especially for low SNRs~\cite{Vallisneri:2007ev}. 
We incorporate Gaussian priors $\Gamma^p_{ii}=1/(\delta P_i)^2$ into the Fisher information matrix,
where $\delta P_i$ represents the maximum allowable variation in the parameter~\cite{Cutler:1994ys, Favata:2013rwa, Favata:2021vhw, Cho:2022cdy}. 
The Fisher matrix approach, with the inclusion of Gaussian priors, has been shown to be consistent with the more computationally intensive MCMC method~\cite{Favata:2021vhw, Cho:2022cdy}.

We select the reference frequencies $f_2 = f(\ell=2)$ to correspond to the time-to-merger $t_c-t$. 
Each harmonic mode enters the detector band at different time (hence different $f_2$). 
The calculations of SNR and localization are performed at several tens of $f_2$ points with adaptive step sizes to ensure a sufficiently smooth evolution.
Our focus is on quantifying how much earlier the eccentric cases can, on average, accumulate a specific SNR and localization compared to the non-eccentric cases.

\section{SNR versus time-to-merger}
\begin{figure*}
\centering
\includegraphics[width=0.9\textwidth]{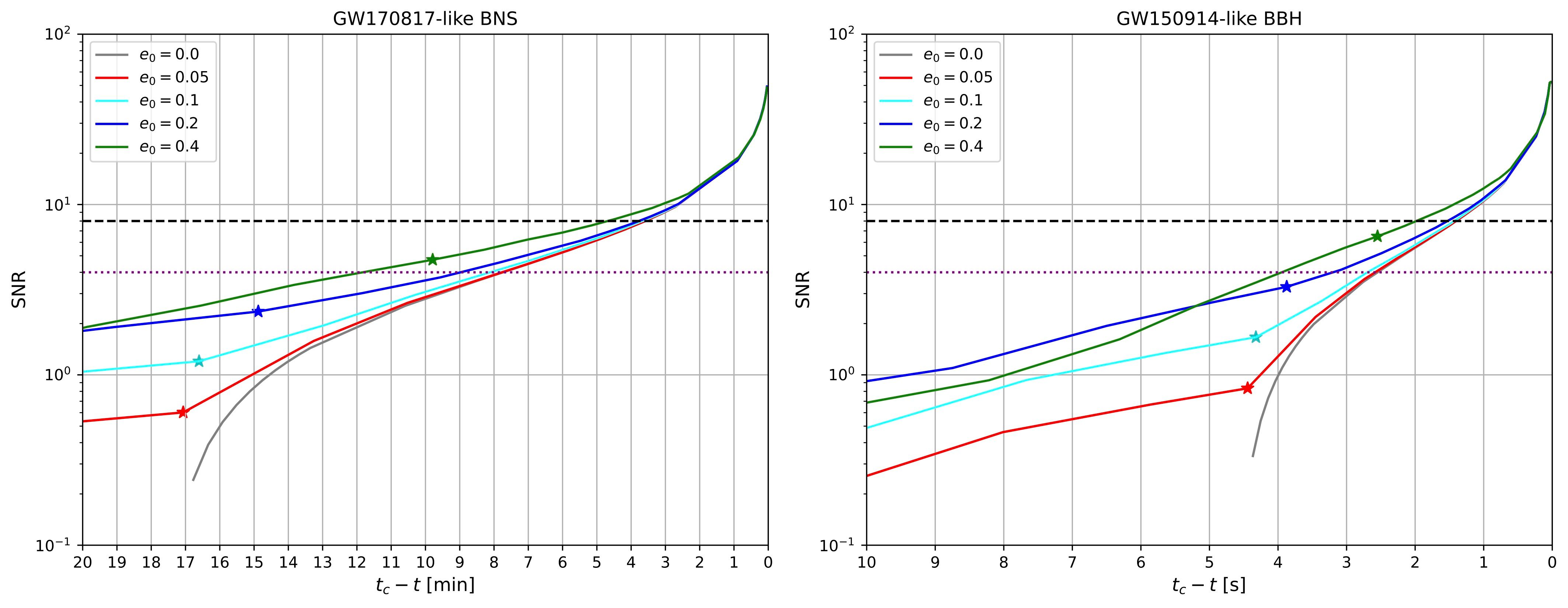}
\caption{The average SNR versus time-to-merger for the typical BNS (left) and BBH (right) event with 2G detector networks HLVKI. The stars indicate the time when the dominant $\ell=2$ mode enters the band at 10 Hz. The two dashed lines indicate the SNR of 4 and 8, respectiviely. We only present the SNR results for the period starting 20 minutes (for BNS) and 10 seconds (for BBH) before the merger.}
\label{fig:snr}
\end{figure*}

We begin calculating the accumulated SNR once the higher mode $\ell=10$ enters the detector band at 10 Hz. 
This allows us to evaluate how much SNR is accumulated before the dominant mode enters the band and to observe the enhancement of the accumulated SNR contributed by higher modes after the dominant mode becomes in-band.
The results for the evolution of average SNR as a function of time-to-merger are shown in Fig.~\ref{fig:snr}.
The left panel shows the results for GW170817-like BNS cases with various eccentricities, while the right panel illustrates the results for GW150914-like BBH cases.
The stars indicate the time when the dominant $\ell=2$ mode enters the band at 10 Hz. As larger eccentricities shorten the inspiral time, the time-to-merger when the dominant mode becomes in-band is smaller for cases with higher eccentricities.

To quantify the time-to-merger gained by eccentricity-induced higher modes compared to non-eccentric cases, 
we set two reference SNR values: $\rho = 4$, corresponding to the lowest SNR in the LVK GW catalogs, and $\rho = 8$, the conventional threshold for confirming a GW event, as indicated by the two horizontal lines.
For the non-eccentric case of GW170817-like BNS, the SNR of 4 is achieved at a time-to-merger of approximately 7.5 minutes, while for the eccentric case with $e_0 = 0.4$, the time-to-merger improves to 12 minutes, representing a 4.5-minute improvement. 
For the threshold SNR of 8, the time-to-merger increases from 3.5 minutes to 5 minutes, yielding a 1.5-minute improvement. However, for smaller eccentricities, the improvement is less significant.
In the typical GW150914-like BBH case, where the inspiral time is relatively short, the eccentric case with $e_0 = 0.4$ achieves an SNR of 4 approximately 1.5 seconds earlier than the non-eccentric case. 
For the threshold SNR of 8, the improvement is only 0.5 seconds.

We observe that the eccentric cases begin to accumulate SNR much earlier than the non-eccentric cases, although we only present the SNR results for the period starting 20 minutes (for BNS) and 10 seconds (for BBH) before the merger. 
Additionally, with larger eccentricities, the SNR accumulates more rapidly due to the increased prominence of higher modes.
An exception is observed for the cases with $e_0 = 0.2$ and $e_0 = 0.4$, where the accumulated SNR for $e_0 = 0.4$ is smaller than that for $e_0 = 0.2$ at earlier times (for BBH cases). 
This behavior arises from two competing effects. On the one hand, a larger eccentricity excites stronger higher harmonic modes, which can in principle increase the accumulated SNR contributed by these modes. On the other hand, a higher eccentricity significantly shortens the inspiral duration. At early times before merger--especially prior to the entry of the dominant mode into the detector band--this leads to fewer higher harmonic modes being present in band and a smaller frequency span covered by each mode (at a fixed time-to-merger). As a result, the accumulated SNR at early times can be lower for $e_0=0.4$ than for $e_0=0.2$. At later times, when more higher harmonics have entered the band, the enhancement due to stronger higher modes dominates, and the SNR for $e_0=0.4$ exceeds that for $e_0=0.2$.
In some cases, $e_0 = 0.2$ represents the optimal scenario. A similar feature was also identified in our previous research.

We also observe that the SNR for all cases converges in the final stage before merger. 
This indicates that the dominant $\ell=2$ mode accumulates SNR very rapidly in the later stages, contributing the majority of the total SNR compared to the higher modes.
A further explanation is as follows: the strength of the subdominant modes is only $\mathcal{O}(10\%)$ of the dominant mode. 
Thus, for both eccentric and non-eccentric cases, the primary contribution to the SNR comes from the $\ell=2$ mode. 
Additionally, since the initial eccentricity is set at 10 Hz,
the remaining eccentricity decreases significantly by the time the system evolves to frequencies where the noise power spectral density is optimal (recall the inverse relation between eccentricity and orbital frequency).
As a result, the higher modes mainly contribute at earlier times when the noise is relatively large. 
Consequently, the SNR for eccentric cases does not show substantial improvement compared to the non-eccentric cases near the merger.

\section{Localization versus time-to-merger}
\begin{figure*}
\centering
\includegraphics[width=0.9\textwidth]{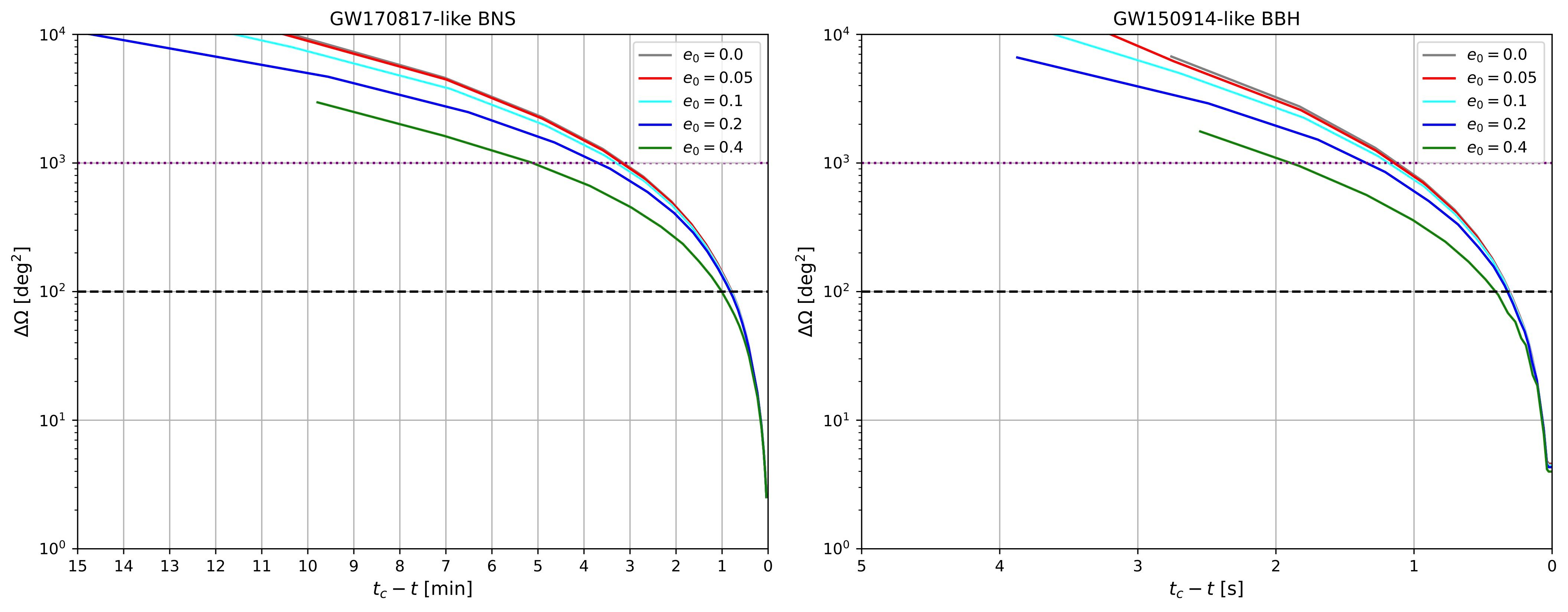}
\caption{The average localization versus time-to-merger for the typical BNS (left) and BBH (right) event with 2G detector networks HLVKI. The two dashed lines indicate the localization of 1000 and 100 $\rm deg^2$, respectiviely. We only present the localization results for the period starting 15 minutes (for BNS) and 5 seconds (for BBH) before the merger.}
\label{fig:loc}
\end{figure*}

Our previous research concluded that localization based solely on higher modes before the dominant $\ell=2$ mode enters the detector at 10 Hz is quite insufficient ($>10^3 \, \rm deg^2$). 
Therefore, we calculate the evolution of localization only after the $\ell=2$ mode has entered the band. 
The results are presented in Fig.~\ref{fig:loc}, where we plot the evolution of localization only when its accuracy $\Delta \Omega < 10^4 \, \rm deg^2$.
For a typical BNS, achieving a localization of $1000 \, \rm deg^2$ requires 3 minutes before merger for the non-eccentric case, 
while the $e_0 = 0.4$ eccentric case achieves the same localization at 5 minutes of time-to-merger, representing a 2-minute improvement.
For a localization of $\Delta \Omega = 100 \, \rm deg^2$, the eccentric case reaches this level 1 minute before merger, compared to about 45 seconds before merger for the non-eccentric case.
In the typical BBH case, a localization of $1000 \, \rm deg^2$ can be achieved 2 seconds before the merger for $e_0 = 0.4$, compared to 1.2 seconds before the merger for $e_0 = 0$. 
Due to the short inspiral time, the improvement in localization to $100 \, \rm deg^2$ from eccentricity-induced higher modes is negligible, around $\mathcal{O}(0.1)$ seconds.
For BBH systems, the inspiral duration in the ground-based detector band is typically only of order seconds. Increasing the eccentricity further shortens this already brief inspiral phase, leaving little time for additional higher harmonic modes to accumulate sufficient SNR to substantially improve sky localization. Consequently, the gain in early-warning time or localization accuracy from eccentricity-induced higher modes is much more limited for BBH systems than for BNS systems, where the inspiral lasts significantly longer. 

We observe that a higher eccentricity results in better localization at early times, while all cases converge at late times just before the merger. This behavior is similar to what was explained for the SNR results.
The final localization near merger is on the order of $\mathcal{O}(1) \, \rm deg^2$ for all BNS and BBH cases. For BNS cases near the merger, eccentricity does not yield an improvement in localization. 
However, for BBH cases, there is a slight improvement, as seen near the merger time in the right panel of Fig.~\ref{fig:loc} (note that $f_{\rm max}$ is truncated at the innermost stable orbit, resulting in a flat evolution beyond that point).
This conclusion is consistent with\cite{Pan:2019anf}, which investigated the impact of eccentricity on source localization when accounting for the full inspiral period, i.e., the localization achieved near merger, in the 2G ground-based detector network.

\section{Conclusion and discussions}
In this paper, we present the first theoretical and detailed investigation of how eccentricity-induced higher harmonic modes enhance the early detection and localization of compact binaries in the 2G detector network HLVKI. 
We find that in eccentric cases, higher harmonic modes, which enter the detector band much earlier than the dominant mode, significantly improve both the SNR and localization accuracy at earlier times compared to non-eccentric cases.
In particular, for a GW170817-like binary neutron star (BNS) with $e_0=0.4$ at 10 Hz, the signal reaches SNR 4 and the detection threshold of SNR 8 approximately 12 and 5 minutes before merger, respectively—gains of 4.5 and 1.5 minutes over the circular case. Localization within $1000 \, (100)\,\rm deg^2$ is achievable 5 (1) minutes before merger, improving by 2 minutes (15 seconds). In the cases of $e_0=0.05$ and 0.1, the improvement is not significant. While the effect is more pronounced for BNS mergers, a modest improvement is observed for BBH cases. 

The search for EM counterparts relies critically on early GW alerts~\cite{Chu:2015jxa, Banerjee:2022gkv}, particularly for BNS mergers, which are the most promising sources of accompanying EM signals. While the duration of detectable GW signals in 2G detectors remains limited compared to 3G and space-based observatories, eccentricity-induced higher modes offer an opportunity to provide earlier alerts for both detection and localization. This extended lead time allows for better preparation of EM telescopes, increasing the likelihood of capturing potential EM counterparts and advancing the prospects for multi-messenger astronomy.

The waveform model adopted in this work is tailored to our specific goal of isolating eccentricity-induced higher harmonic modes and assessing their impact on early detection and localization. Our analysis requires tracking, at fixed pre-merger times, which orbital harmonics have entered the detector band and the frequency range spanned by each harmonic, which in turn necessitates a frequency-domain waveform in which individual harmonics can be explicitly separated and independently assigned detector response functions. EccentricFD provides this capability and is therefore well suited to our framework. While more recent eccentric waveform models including merger and ringdown phases are available, they are formulated in the time domain and do not allow for a straightforward extraction of individual eccentricity-induced harmonics. Moreover, since our focus is on early warning during the inspiral--even at epochs when the dominant $\ell=2$ mode may have not yet entered the detector band--only the inspiral phase is relevant by construction, and neglecting merger–ringdown phases does not affect our early-warning conclusions.

In our simulations of typical GW events, we fixed their distances to match the median values of the true events reported by LVK. Our primary objective is to examine the newly proposed idea, and these representative events with their true distances serve as sufficient examples for demonstration. However, our methodology can be readily extended to other distances, following the approximate relationships in our calculations: $\rho \sim 1/d_L$ and $\Delta\Omega \sim 1/\rho^2 \sim d_L^2$.

The occurrence rate of eccentric binaries remains an active area of research~\cite{Ye:2019xvf,Samsing:2013kua,Samsing:2017xmd,Samsing:2017rat}, with estimates highly dependent on the uncertain astrophysical formation channels of these compact binaries. 
The astrophysical occurrence rate of BNS mergers with measurable eccentricity in the ground-based detector band remains highly uncertain and is likely small. Current cluster simulations indicate that globular clusters contribute only a minor fraction of the overall BNS merger rate; for example,~\cite{Ye:2019xvf} estimates a local BNS or NSBH merger rate from globular clusters of $\sim 0.02\,{\rm Gpc^{-3}\,yr^{-1}}$, corresponding to a sub-percent (and plausibly much smaller) fraction of the total BNS population. At the same time, recent observational studies suggest that highly eccentric compact-binary mergers do occur in nature. Beyond GW190521, several BBH mergers with inferred eccentricities of order $\sim 0.3$ have been reported~\cite{Gupte:2024jfe,Planas:2025jny}. More intriguingly, the NSBH event GW200105 has been argued to exhibit non-negligible eccentricity~\cite{Morras:2025xfu,Kacanja:2025kpr,Jan:2025fps,Phukon:2025cky}, with $e_0\gtrsim 0.1$ at 20 Hz, corresponding to an even larger eccentricity at 10 Hz. These findings motivate continued exploration of eccentric neutron-star–containing binaries, despite their expected rarity. In this context, the choice of $e_0=0.4$ in this work should be interpreted as an upper-envelope case, within the validated domain of the waveform model, used to quantify the maximum potential early-warning gains enabled by eccentricity-induced higher harmonics. Our results show that for small eccentricities ($e_0\le 0.1$) the improvement in warning time is modest, whereas eccentricities of $e_0\sim 0.2\text{–}0.4$ are required to achieve minute-scale gains for GW170817-like BNS systems. The expected number of such detections scales as $N_{\rm ecc}\approx f_{\rm ecc}N_{\rm BNS}$, where $f_{\rm ecc}$ is the fraction of eccentric systems. While $f_{\rm ecc}$ may be small, even a single confidently identified eccentric BNS merger would be of high scientific value, particularly for multi-messenger observations where additional minutes of advance warning can be decisive.

Since our goal is to estimate the average SNR and sky localization accuracy over 1000 realizations of angular parameters, we adopt the Fisher-matrix formalism as a computationally efficient, theoretically well-motivated approximation. While informative priors are included, it is well known that Fisher-matrix estimates can overestimate parameter estimation accuracy in the low-SNR regime, particularly near detection threshold. As a result, the absolute localization areas reported here should be interpreted as optimistic estimates in this regime. However, because our analysis focuses on relative improvements in localization enabled by eccentricity-induced higher harmonics--rather than on absolute localization precision--we expect our qualitative conclusions to be robust. A more accurate assessment of sky localization near threshold would require full Bayesian parameter estimation using, for example, Markov Chain Monte Carlo methods, which we leave for future work.

\begin{acknowledgments}This work is supported by the National Natural Science Foundation of China Grants No. 12575063.
Part of the numerical calculations in this paper have been done on the supercomputing system in the Supercomputing Center of Wuhan University.
R.G.C is supported by the National Key Research and Development Program of China Grant No. 2020YFC2201502 and 2021YFA0718304 and by National Natural Science Foundation of China Grants No. 11821505, No. 11991052, No. 11947302, No. 12235019.
Z.C is supported in part by the National Key Research and Development Program of China Grant No. 2021YFC2203001, in part by ``the Fundamental Research Funds for the Central Universities''.
H.M.L is supported by the National Research Foundation of Korea 2021M3F7A1082056.
\end{acknowledgments}

\bibliography{ref}

\end{document}